\documentclass[pre,aps,twocolumn,showpacs]{revtex4}
\usepackage{epsfig}
\usepackage{graphicx}
\usepackage{textcomp}

\begin{document}
\title{Spectral theory of metastability and extinction in a branching-annihilation reaction}
\author{Michael Assaf and Baruch Meerson}
\affiliation{Racah Institute of Physics, Hebrew University of Jerusalem,
Jerusalem 91904, Israel} \pacs{05.40.-a, 02.50.Ey, 87.23.Cc, 82.20.-w}

\begin{abstract}
We apply the spectral method, recently developed by the authors, to calculate
the statistics of a reaction-limited multi-step birth-death process, or chemical
reaction, that includes as elementary steps branching $A
\to\hspace{-2.8mm}\hspace{2mm} \,\,2A$ and annihilation $2A
\to\hspace{-2.8mm}\hspace{2mm}\,\, \emptyset$. The spectral method employs the
generating function technique in conjunction with the Sturm-Liouville theory of
linear differential operators. We focus on the limit when the branching rate is
much higher than the annihilation rate, and obtain accurate analytical results
for the complete probability distribution (including large deviations) of the
metastable long-lived state, and for the extinction time statistics. The
analytical results are in very good agreement with numerical calculations.
Furthermore, we use this example to settle the issue of the ``lacking" boundary
condition in the spectral formulation.
\end{abstract}

\maketitle

\section{Introduction}
Statistics of rare events, or large deviations, in chemical
reactions and systems of birth-death type have attracted a great
deal of interest in many areas of science including physics,
chemistry, astrochemistry, epidemiology, population biology, cell
biochemistry, \textit{etc.} \cite{gardiner,vankampen,
bartlett,karlin,karlin2,hanski,delb,McQuarrie,Darroch,brau,malek,oppenheim,gillespie,Biham}.
Large deviations become of vital importance when discrete
(non-continuum) nature of a population of ``particles" (molecules,
bacteria, cells, animals or even humans) drives it to extinction. A
standard way of putting the discreteness of particles into theory is
the master equation \cite{gardiner,vankampen} which describes the
evolution of the probability of having a certain number of particles
of each type at time $t$. The master equation is rarely soluble
analytically, and various approximations are in use
\cite{gardiner,vankampen}. One widely used approximation is the
Fokker-Planck equation which usually gives accurate results in the
regions around the peaks of the probability distribution, but fails
in its description of large deviations, that is the distribution
tails \cite{gaveau,kamenev,Sander}. Not much is known beyond the
Fokker-Planck description. In some particular cases (especially, for
single-step birth-death processes) complete statistics, including
large deviations, were determined by applying various approximations
directly to the pertinent master equation \cite{brau,
oppenheim,Biham,Sander,turner,dykman,nasell1,laurenzi,nasell2,ovaskainen}.
A different group of approaches employs the generating function
formalism \cite{gardiner,karlin2,McQuarrie}, see below. Here the
master equation is transformed into a linear partial differential
equation (PDE) for the generating function, and this PDE is
analyzed/solved by various techniques such as the method of second
quantization \cite{Doi,Peliti,Mattis} or the more recent
time-dependent WKB approximation \cite{kamenev,escudero}. Recently,
we combined the generating function technique with the
Sturm-Liouville theory of linear differential operators and
developed a \textit{spectral theory} of rare events
\cite{assaf,assaf1}. In this theory the problem of computing the
complete statistics of (not necessarily single-step) birth-death
systems reduces to solving an eigenvalue problem for a linear
differential operator, the coefficients of which are determined by
the reaction rates.

In this paper we apply the spectral method to the paradigmatic
problem  of branching $A+X\to 2X$ and annihilation $X+X\to E$, where
$A$ and $E$ are fixed. This multi-step single-species birth-death
process describes, for example, chemical oxidation reactions
\cite{gates,turner}. If the branching rate is much higher than the
annihilation rate (the case we will be mostly interested in
throughout the paper), a long-lived metastable, or quasi-stationary,
state exists where the two processes (almost) balance each other.
Still, this long-lived state slowly decays with time, because a
sufficiently large fluctuation ultimately brings the system into the
absorbing state of no particles from which there is a zero
probability of exiting. In this type of problems one is interested
in the extinction time statistics and in the complete probability
distribution, including large deviations, of the quasi-stationary
state (formally defined as the limiting distribution conditioned on
non-extinction). Turner and Malek-Mansour \cite{turner} calculated
the mean extinction time in this system by solving a recursion
equation for the extinction probability. More recently, Elgart and
Kamenev \cite{kamenev} revisited  this problem in the light of their
time-dependent WKB approximation for the generating function. Their
insightful method readily yields an estimate of the mean extinction
time, but only up to a (significant) pre-exponential factor. The
quasi-stationary distribution for this system has not been
previously found, and calculating it will be our major objective. In
the language of the spectral theory, the mean extinction time
represents the inverse eigenvalue of the ground state, while the
quasi-stationary distribution is derivable from the ground state
eigenfunction.

The paradigmatic branching-annihilation problem, considered in this
paper, has an additional value, as it helps settle one unresolved
issue of the spectral theory. In the previous works
\cite{assaf,assaf1} we considered reactions that conserve parity of
the particles.  The parity conservation provides an additional
boundary condition for the PDE for the generating function which
ensures a closed formulation of the problem already at the stage of
the time-dependent PDE. The branching-annihilation process,
considered in the present work, does \textit{not} conserve parity.
As we will show, the ``lacking" boundary condition emerges here (and
in a host of other problems of this type) only at the stage of the
Sturm-Liouville theory.

Here is how we organize the rest of the paper. In Section II we apply the
spectral method and reduce the governing master equations to a proper
Sturm-Liouville problem. In Section III we employ a matched asymptotic expansion
to approximately calculate the ground-state eigenvalue and eigenfunction and
obtain the long-time asymptotics of the generating function. This asymptotics is
used in Section IV to extract the quasi-stationary probability distribution and
compare it with our numerical results. In Section V we calculate the mean
extinction time and extinction probability distribution and compare these
results with the previous work and with our numerics. Some final comments are
presented in Section VI.

\section{Generating function and Spectral formulation}
We consider the branching and annihilation reactions $A
\to\hspace{-4.3mm}^{\lambda}\hspace{2mm} 2A$ and  $2A
\to\hspace{-4.3mm}^{\mu}\hspace{2mm} \emptyset$, where $\mu,\lambda>0$ are the
rate constants. The (mean-field) rate equation for the number of particles
$n(t)$, $dn/dt=\lambda n-\mu n^2$ predicts a nontrivial attracting steady state
$n_s=\lambda/\mu\equiv \Omega$. Fluctuations invalidate this mean-field result
due to the existence of an absorbing state at $n=0$. However, when $\Omega\gg
1$, there exists a long-lived fluctuating \textit{metastable} (or
quasi-stationary) state, which \textit{slowly} decays in time, implying a slow
growth of the extinction probability. The statistics of this quasi-stationary
state and of the extinction times are in the focus of our attention here.

The master equation for the probability $P_n(t)$ to find $n$ particles at time
$t$ can be written  as
\begin{eqnarray}
\label{mastereq}
\frac{d}{dt}{P}_{n}(t)&=&\frac{\mu}{2}\left[(n+2)(n+1)
P_{n+2}(t)-n(n-1)P_{n}(t)\right]\nonumber\\
&+&\lambda\left[(n-1)P_{n-1}(t)-nP_{n}(t)\right]\,,\;\;\;n\ge1\,,\nonumber\\
\frac{d}{dt}{P}_{0}(t)&=& \mu P_2(t)\,.
\end{eqnarray}
We introduce the generating function \cite{McQuarrie,gardiner,vankampen}
\begin{equation}\label{generatingfun}
G(x,t)=\sum_{n=0}^{\infty}x^{n}P_{n}(t)\,,
\end{equation}
where $x$ is an auxiliary variable. Once $G(x,t)$ is known, the probabilities
$P_n(t)$ can be recovered from the Taylor expansion:
\begin{equation}\label{probformula}
\left.P_{n}(t)=\frac{1}{n!}\frac{\partial^{n}G(x,t)}{\partial
x^{n}}\right|_{x=0}\,.
\end{equation}
By virtue of Eqs.~(\ref{generatingfun}) and (\ref{probformula}), $G(x,t)$ must
be analytical, at all times, at $x=0$. Equations~(\ref{mastereq}) and
(\ref{generatingfun}) yield a single PDE for $G(x,t)$ \cite{kamenev}:
\begin{equation}\label{partdiffeq}
\frac{\partial G}{\partial t} =
\frac{\mu}{2}(1-x^{2})\frac{\partial^{2} G}{\partial x^{2}}+\lambda
x(x-1)\frac{\partial G}{\partial x}\,.
\end{equation}
Conservation of probability yields one (universal) boundary condition for this
parabolic PDE: $G(1,t)=1$ \cite{math}. What is the second boundary condition?
Note that $G(x=-1,t)$ must be bounded at all times, as it is equal to the
difference between the sum of the probabilities to have an even number of
particles and the sum of the probabilities to have an odd number of particles.
Now, the \textit{steady-state} solution of Eq.~(\ref{partdiffeq}),
$G_{st}(x)=G(x,t\to \infty)$, that obeys the equation
\begin{equation}\label{steady}
\frac{\mu}{2}(1-x^{2})\frac{d^{2} G_{st}}{d x^{2}}+\lambda
x(x-1)\frac{d G_{st}}{d x}=0\,.
\end{equation}
must also be bounded at $x=-1$. Then Eq.~(\ref{steady}) immediately yields a
second boundary condition: $G_{st}^{\prime}(x)|_{x=-1}=0$, where the prime
stands for the $x$-derivative. Combined with $G_{st}(1)=1$, this condition
selects the steady state solution $G_{st}(x)=1$ describing an empty state.

Now let us introduce a new function $g(x,t)=G(x,t)-G_{st}(x)=G(x,t)-1$ [which
obeys Eq.~(\ref{partdiffeq}) with a homogenous boundary condition $g(x= 1,t)=0$
and is bounded at $x=-1$], and look for separable solutions,
$g_k(x,t)=e^{-\gamma_k t}\varphi_k(x)$. We obtain
\begin{equation}\label{orddiffeq}
(1-x^{2})\varphi_k^{\prime\prime}(x)+2 \Omega x(x-1)\varphi_k^{\prime}(x)+2
E_k\varphi_k(x)=0\,,
\end{equation}
where $E_k=\gamma_k/\mu$. One boundary condition is of course $\varphi_k(1)=0$.
The second boundary condition comes from the demand that $\varphi_k(x)$ be
bounded at $x =-1$. Then Eq.~(\ref{orddiffeq}) yields a homogenous boundary
condition
\begin{equation}\label{second}
2\Omega\varphi_{k}^{\prime}(-1)+E_{k}\varphi_{k}(-1)=0\,,
\end{equation}
for each $k=1,2,\dots$. Notice that the eigenvalue $E_k$ enters the boundary
condition. Rewriting Eq. (\ref{orddiffeq}) in a self-adjoint form,
\begin{equation}\label{schroed}
\left[\varphi_k^{\prime}(x)\exp(-2\Omega x)(1+x)^{2\Omega}\right]^{\prime}+E_k
w(x) \varphi_k(x)=0\,,
\end{equation}
with the weight function
\begin{equation}\label{weight}
w(x) = \frac{2 e^{-2\Omega x}(1+x)^{2\Omega}}{1-x^2}\,,
\end{equation}
we arrive at an eigenvalue problem of the Sturm-Liouville theory \cite{Arfken}.
Once the complete set of orthogonal eigenfunctions $\varphi_{k}(x)$ and the
respective real eigenvalues $E_{k}$, $k=1,2,\dots$, are calculated, one can
write the exact solution of the time-dependent problem for $G(x,t)$:
\begin{equation}\label{genseries}
G(x,t)= 1+\sum_{k=1}^{\infty} a_k \varphi_{k}(x) e^{-\mu
E_{k}t}\,\,,
\end{equation}
where the amplitudes $a_k$ are given by
\begin{equation}\label{an}
a_{k}=\frac{\int_{-1}^{1}[G(x,t=0)-1]\varphi_{k}(x)w(x)dx}{\int_{-1}^{1}\varphi_{k}^2(x)w(x)dx}.
\end{equation}
As all $E_k$ are positive, Eq. (\ref{genseries}) describes \textit{decay} of
initially populated states $k=1, 2, \dots$,  so the system ultimately approaches
the empty state $G(x,t\to \infty)=1$. Being mostly interested in the case of
$\Omega\gg 1$, we note that while the eigenvalues of the ``excited states"
$E_{2}, E_{3},\dots$ scale like ${\cal O}(\Omega)\gg 1$ \cite{kamenev1}, the
``ground state" eigenvalue $E_{1}$ is \textit{exponentially} small
\cite{turner}. Therefore, at sufficiently long times, $\mu \Omega t=\lambda t\gg
1$, the contribution from the excited states to $G(x,t)$ becomes negligible, and
we can write
\begin{equation}\label{genseriesapprox}
G(x,t)= 1+a_1 \,\varphi_{1}(x)\,e^{-\mu E_{1}t}.
\end{equation}
So we need to calculate the ground-state eigenvalue $E_1$, the eigenfunction
$\varphi_1(x)$, and the amplitude $a_1$. (Actually, the eigenvalue $E_1$ was
calculated earlier \cite{turner}, but we will rederive it here.) Note that, as
$E_1$ is exponentially small, the boundary condition (\ref{second}) for the
ground state reduces, up to an exponentially small correction, to
\begin{equation}\label{second1}
    \varphi_1^{\prime}(-1)=0\,.
\end{equation}

\section{Ground state calculations}
Throughout the rest of the paper we assume $\Omega\gg1$. As $E_1$ is
exponentially small, the last term in Eq.~(\ref{orddiffeq}) is important only in
a narrow boundary layer near $x=1$, and we can solve Eq.~(\ref{orddiffeq}) for
$\varphi_1(x)\equiv \varphi(x)$ by using a matched asymptotic expansion, see
\textit{e.g.} Ref. \cite{orszag}. In the ``bulk" region $-1\leq x<1$ we can
treat the last term in Eq. (\ref{orddiffeq}) perturbatively. In the zeroth order
we put $E_1=0$ and arrive at the \textit{steady state} equation
$(1+x)\varphi^{\prime\prime}(x)-2 \Omega x\varphi^{\prime}(x)=0\,,$ whose
(arbitrarily normalized) solution, bounded at $x=-1$, is $\varphi_b^{(0)}(x)=1$.
Now we put $\varphi_{b}(x)=1+\delta\varphi_{b}(x)$, where $\delta\varphi_{b}(x)
\ll 1$, and obtain in the first order
\begin{equation}\label{orddiff1}
\left[\delta\varphi_{b}^{\prime}(x)e^{-2\Omega x}
(1+x)^{2\Omega}\right]^{\prime}=-\frac{2E_{1} e^{-2\Omega
x}}{1-x^{2}}(1+x)^{2\Omega}\,.
\end{equation}
Solving this equation, we obtain the bounded solution for
$\varphi_b(x)$:
\begin{equation}\label{phibulk}
\varphi_{b}(x)=1-2E_{1}\int_{0}^{x}\frac{e^{2\Omega
s}ds}{(1+s)^{2\Omega}}\int_{-1}^{s}\frac{(1+r)^{2\Omega}e^{-2\Omega
r}}{1-r^2}dr\,.
\end{equation}
This solution, that obeys the boundary condition (\ref{second}), is almost
constant in the entire region $-1\leq x< 1$ except in the boundary layer near
$x=1$ (to be defined later on). To find the probabilities $P_n(t)$, we will need
to calculate the derivatives of $\varphi_b(x)$ at $x=0$. As long as $1-x\gg
1/\Omega$, we can neglect the $r^2$ term in the denominator of the inner
integral in Eq. (\ref{phibulk}) and obtain
\begin{eqnarray}
\varphi_{b}(x)&\simeq&1-2E_{1}\int_{0}^{x}\frac{e^{2\Omega
s}ds}{(1+s)^{2\Omega}}\int_{-1}^{s}(1+r)^{2\Omega}e^{-2\Omega
r}dr\nonumber\\
&=&
1-\frac{E_{1}}{\Omega}\left(\frac{e}{2\Omega}\right)^{2\Omega}\int_{0}^{x}\frac{e^{2\Omega
s}}{(1+s)^{2\Omega}}\left\{\Gamma[2\Omega+1]\right.\nonumber\\
&-&\left.\Gamma[2\Omega+1,2\Omega(1+s)]\right\}ds,
\label{phibulkapp}
\end{eqnarray}
where $\Gamma(\alpha,z)=\int_z^{\infty}s^{\alpha-1}e^{-s}ds\;$ is
the incomplete Gamma function \cite{Abramowitz}. Using the expansion
\cite{Gradstein}
\begin{equation}\label{gammaexpansion}
\Gamma(\alpha)-\Gamma(\alpha,z)=\sum_{j=0}^{\infty}\frac{(-1)^j
z^{\alpha+j}}{j! (\alpha+j)}\,,
\end{equation}
we can evaluate the integral in Eq.~(\ref{phibulkapp}) and obtain
\begin{eqnarray}
\varphi_{b}(x)\simeq
1-\frac{E_{1}}{2\Omega^2}\sum_{j=0}^{\infty}\frac{\Gamma[2\!+\!j,-2\Omega]-\Gamma[2\!+\!j,-2\Omega(1\!+\!x)]}{j!(2\Omega+j+1)}\,.\nonumber\\
\label{phibulkapp1}
\end{eqnarray}
One can check that the perturbative solution in the bulk is valid [that is,
$\delta\varphi_{b}(x)\ll 1$] as long as $1-x\gg 1/\Omega$.

In the boundary layer $1-x\ll 1$ we can disregard, in Eq.~(\ref{orddiffeq}), the
(exponentially small) last term, and arrive at the same equation as before:
$(1+x)\varphi^{\prime\prime}(x)-2\Omega x\varphi^{\prime}(x) =0\,.$ The solution
obeying the required boundary condition at $x=1$ is
\begin{eqnarray}
\varphi_{bl}(x)&=& const \times \int_{1}^{x}e^{2\Omega s}(1+s)^{-2\Omega}ds\nonumber\\
&\simeq& C\left[1-e^{-2\Omega(1-\ln 2)}\frac{e^{2\Omega
x}}{(1+x)^{2\Omega}}\right]\,, \label{phiwallright}
\end{eqnarray}
where $C$ is a yet unknown constant. To find $E_{1}$ and $C$ we can match the
asymptotes of the bulk and the boundary-layer solutions in the common region of
their validity $1/\Omega \ll 1-x\ll 1$. Let us return to the first line of Eq.
(\ref{phibulkapp}) and evaluate $\varphi_b(x)$ in this region. The inner
integral receives the largest contribution from the vicinity of $r=0$, while the
outer integral receives the largest contribution from the vicinity of $s=x$.
Therefore, we can extend the upper limit of the inner integral to infinity and
obtain (see Appendix A)
\begin{eqnarray}
\varphi_{b}(x)&\simeq&1-2E_{1}\int_{0}^{x}\frac{e^{2\Omega
s}ds}{(1+s)^{2\Omega}}\int_{-1}^{\infty}(1+r)^{2\Omega}e^{-2\Omega
r}dr \nonumber\\
&\simeq& 1-
\frac{2E_{1}\sqrt{\pi}}{\sqrt{\Omega}}\int_{0}^{x}\frac{e^{2\Omega
s}}{(1+s)^{2\Omega}}ds \nonumber\\
&\simeq& 1- \frac{2E_{1}\sqrt{\pi}}{\Omega^{3/2}}\frac{e^{2\Omega
x}}{(1+x)^{2\Omega}}\,, \label{phibulkapprox}
\end{eqnarray}
Now, by matching Eqs. (\ref{phiwallright}) and
(\ref{phibulkapprox}), we obtain
\begin{equation}\label{results}
E_{1}=\frac{\Omega^{3/2}}{2\sqrt{\pi}}e^{-2\Omega(1-\ln 2)
}\;\;\;\mbox{and}\;\;\;C=1\;\,.
\end{equation}
One can see that the ground state eigenvalue $E_{1}$ is exponentially small in
$\Omega$. Equation (\ref{results}) yields the mean extinction time $(\mu
E_1)^{-1}$ (see Section V) which coincides with that obtained, by a different
method, by Turner and Malek-Mansour \cite{turner}.

Equations (\ref{phibulk}) and (\ref{phiwallright}) yield the ground state
eigenfunction:
\begin{equation}\label{eigenfunction}
\varphi_1(x)\simeq\left\{\begin{array}{ll}
\varphi_{b}(x)\;\;\;\;\;\mbox{for}\; 1-x\gg1/\Omega\,,\nonumber\\
\\
\varphi_{bl}(x)\;\;\;\;\mbox{for} \; 1-x\ll 1\,.\end{array}\right.
\end{equation}
Now we use Eq.~(\ref{an}) to calculate the amplitude $a_{1}$
entering Eq.~(\ref{genseriesapprox}). Let the initial number of
particles be $n_0$, so $G(x,t=0)=x^{n_0}$. Evaluating the integrals,
we notice that (i) the main contributions come from the bulk region
$1-x\gg 1/\Omega$, and (ii) it suffices to take the eigenfunction
$\varphi_b(x)$ in the zeroth order: $\varphi_b^{(0)}(x)\simeq 1$.
Furthermore, when $n_0\gg1$, the term $x^{n_0}$ under the integral
in the numerator is negligible compared to $1$. So, for $n_0\gg1$,
the numerator and denominator are approximately equal to each other
up to a minus sign. Therefore, $a_{1}\simeq -1$ (and independent of
$n_0$) which completes our solution (\ref{genseriesapprox}) for
times $\mu t\gg \Omega^{-1}$.

\section{Statistics of the quasi-stationary state and its decay}
\begin{figure}[ht]
\includegraphics[width=7.5cm,clip=]{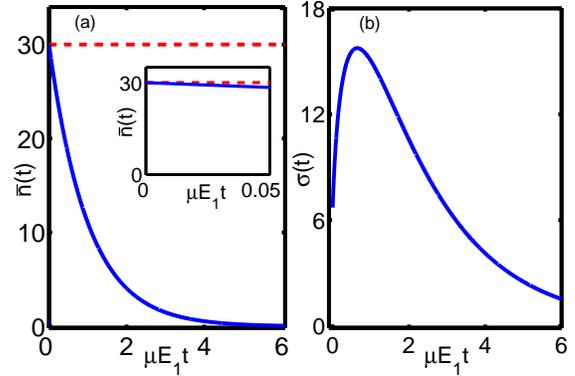}
\caption{(Color online) (a) The average number of particles as a
function of time at $\mu t \gg 1/\Omega$ as described by Eq.
(\ref{nbartime}) (the solid line) compared with the prediction from
the rate equation $\bar{n}(t)\simeq \Omega$ (the dashed line), for
$\Omega=30$. Inset shows a blowup at intermediate times
$\Omega^{-1}\ll\mu t\ll E_{1}^{-1}$ where the curves almost
coincide. (b) The standard deviation from Eq.~(\ref{nvartime})
versus time for the same $\Omega$.} \label{nbarplot}
\end{figure}

What is the average number of particles $\bar{n}(t)$ and the standard deviation
$\sigma(t)$ at times $\mu t\gg\Omega^{-1}$? Using Eq.~(\ref{generatingfun}) and
Eq.~(\ref{genseriesapprox}) with $a_1=-1$, we obtain
\begin{equation}\label{nbartime}
\bar{n}(t)=\sum_{n=0}^{\infty}nP_n(t)=\left.\partial_x G\right|_{x=1}= \Omega
e^{-\mu E_1 t}\,.
\end{equation}
Furthermore,
\begin{eqnarray}
\sigma^2(t)&=&\bar{n^2}-\bar{n}^2=
\sum_{n=0}^{\infty}n^2P_n(t)-\left(\sum_{n=0}^{\infty}nP_n(t)\right)^2\nonumber\\
&=&\left. \left[\partial_{xx}^2 G+\partial_x G -(\partial_x
G)^2\right]\right|_{x=1} \nonumber\\
&=& \left[\frac{3 \Omega}{2}+\Omega^2 \left(1-e^{-\mu E_1
t}\right)\right]e^{-\mu E_1 t}\,, \label{nvartime}
\end{eqnarray}
where we have used for $\varphi_1(x)$ its boundary layer asymptote
$\varphi_{bl}(x)$ (\ref{phiwallright}) with $C=1$. At \textit{intermediate}
times $\Omega^{-1}\ll\mu t\ll E_{1}^{-1}$ one obtains a weakly fluctuating
quasi-stationary (metastable) state. Here the average number of particles,
\begin{equation}\label{nbar}
\bar{n}\simeq \Omega\,,
\end{equation}
coincides with the attracting point of the mean field theory, while the standard
deviation,
\begin{equation}\label{nvar}
\sigma\simeq \left(\frac{3 \Omega}{2}\right)^{1/2}\,,
\end{equation}
coincides with that obtained from the Fokker-Planck approximation, see Appendix
B. Note that $\sigma(t)$ from Eq.~(\ref{nvartime}) is a non-monotonic function.
This stems from the fact that the quasi-stationary probability distribution
around $n\simeq\Omega$ decays in time, whereas the extinction probability
$P_0(t)$ grows. At times $\mu E_1 t\ll 1$ the standard deviation $\sigma\simeq
\sqrt{3\Omega/2}$ corresponds to the unimodal quasi-stationary distribution
around $n\simeq\Omega$, whereas at $\mu E_1 t\gg 1$, $\sigma\to 0$ corresponds
to the unimodal Kronecker delta distribution at $n=0$. At intermediate times,
$\mu E_1 t\simeq 1$, the distribution is distinctly bi-modal. The maximum
standard deviation $\sigma_{max}\simeq\Omega/2$ is obtained for $e^{-\mu E_1
t}\simeq 1/2$. Figure \ref{nbarplot} shows the $\bar{n}(t)$ and $\sigma(t)$
dependences.

Let us now proceed to calculating the complete probability distribution
$P_{n}(t)$ of the (slowly decaying) quasi-stationary state, conditional on
non-extinction. For $n=0$ we obtain
\begin{equation}\label{p0}
P_0(t)=G(x=0,t)= 1 - e^{-\mu E_{1} t}
\end{equation}
which, at $\mu E_{1} t\ll 1$, is much less than unity. For $n\ge 1$
Eqs.~(\ref{probformula}) and (\ref{genseriesapprox}) yield
\begin{eqnarray}
P_{n}(t)&=&\!\left.-\,\frac{1}{n!}\frac{d^{n}\varphi_{b}(x)}{d
x^{n}}\right|_{x=0}\,e^{-\mu E_{1} t}, \label{prob}
\end{eqnarray}
where $\varphi_{b}(x)$ should be taken from Eq. (\ref{phibulkapp}). After some
algebra (see Appendix C), we obtain, for $n\geq 1$,
\begin{equation}\label{probexact}
P_{n}(t)\!=
\!\frac{2E_{1}}{n}\!\frac{(2\Omega)^{n\!-\!1}e^{2\Omega}\Gamma(2\Omega)}
{\Gamma(2\Omega+n)}\,_1\!
F\!_1(2\Omega,n\!+\!2\Omega,-\!2\Omega)e^{-\mu E_1 t},
\end{equation}
where $_1\! F\!_1(a,b,x)$ is the Kummer confluent hypergeometric function
\cite{Abramowitz}. To avoid excess of accuracy, we need to find the large
$\Omega$ asymptotics of Eq. (\ref{probexact}). To that aim we use the identity
\cite{Abramowitz}
\begin{eqnarray}
\!\!\!\!\!\!\!\!\!\!\!\!_1\!F\!_1(2\Omega,n\!+
\!2\Omega,-\!2\Omega)&=&\frac{\Gamma(n+2\Omega)}
{\Gamma(2\Omega)\Gamma(n)}\nonumber\\
&\times&\int_0^1 e^{-2\Omega s}s^{2\Omega-1}(1\!-\!s)^{n\!-\!1}ds \label{p1}
\end{eqnarray}
and consider separately two cases: $n\gg 1$ and $n={\cal O}(1)$.

For $n\gg 1$, the integral in Eq. (\ref{p1}) can be evaluated by the saddle
point method \cite{orszag}. Denoting $\Phi(s)=-2\Omega s+2\Omega\ln(s)+n\ln
(1-s)$ we obtain
\begin{eqnarray}
P_{n}(t)&\simeq&\frac{2E_{1}}{n!}\frac{\sqrt{2\pi}
(2\Omega)^{n-1}e^{2\Omega}}{\sqrt{2\Omega(1-s_*)^2+ns_*^2}}\nonumber\\
&\times&e^{-2\Omega\left[s_*-\ln (s_*)\right]+n\ln(1-s_*)-\mu E_1
t}\,, \label{probapprox}
\end{eqnarray}
where $s_*=1+q-(q^2+2q)^{1/2}$ is the solution of the saddle point equation
$\Phi^{\prime}(s)=0$, and $q=n/(4\Omega)$. Equation~(\ref{probapprox}) can be
simplified in three limiting cases. In the high-$n$ tail, $n\gg \Omega\gg 1$, we
have $s_*\simeq 2\Omega/n\ll 1$, and
\begin{equation}\label{righttail}
P_{n}(t)\simeq\frac{2^{2\Omega-3/2}}{\sqrt{\pi
n}}\left(\frac{2\Omega}{n}\right)^{n+2\Omega}e^{n-2\Omega-\frac{6\Omega^2}{n}-\mu
E_1 t}\,.
\end{equation}
In the low-$n$ tail, $1\ll n\ll\Omega$, we have $s_*\simeq
1-\sqrt{n/(2\Omega)}$, and
\begin{equation}\label{lefttail}
P_{n}(t)\simeq\frac{2^{2\Omega-2}}{\sqrt{\pi
n}}\left(\frac{2\Omega}{n}\right)^{n/2+1/2}e^{n/2-2\Omega-\mu E_1
t}\,.
\end{equation}
Finally, for $|n-\Omega|\ll \Omega$, $s_*\simeq 1/2-(n-\Omega)/(6\Omega)$, and
we obtain
\begin{equation}\label{centerprob}
P_{n}(t)\simeq (3\pi\Omega)^{-1/2}\,e^{-\frac{(n-\Omega)^2}{3\Omega}-\mu E_1
t}\,.
\end{equation}
For $\mu E_1 t \ll 1$ this result describes a normal distribution with mean
$\Omega$ and variance $3\Omega/2$, in agreement with Eqs.~(\ref{nbar}) and
(\ref{nvar}) and with the predictions from the Fokker-Planck equation, see
Appendix B.

Now we turn to the case of $n={\cal O}(1)$. Then it is always $n\ll \Omega$.
Here it is convenient to rewrite the integral in Eq.~(\ref{p1}) as
\begin{equation}\label{p1a}
    \int_0^1 e^{\Psi(s)} \,s^{-1} (1-s)^{n-1}\,ds\,,
\end{equation}
where $\Psi(s)=2\Omega \,(\ln s -s)$. The function $\Psi(s)$ has its maximum
exactly at $s=1$, the upper integration limit. The largest contribution to the
integral comes from the small region ${\cal O}(1/\sqrt{\Omega})$ near $s=1$.
Therefore, it suffices to expand $\Psi(s)$ up to the second order in $(s-1)^2$,
replace the factor $s^{-1}$ by $1$ and extend the lower integration limit to
$-\infty$. The result is
\begin{equation}
e^{-2\Omega}\int_{-\infty}^{1} e^{-\Omega (s-1)^2} (1-s)^{n-1}ds
=\frac{e^{-2\Omega}\,\Gamma\left(n/2\right)}{2 \,\Omega^{n/2}}\,. \label{p2}
\end{equation}
Therefore, for $n={\cal O}(1)$, we obtain
\begin{equation}\label{farlefttail}
P_{n}(t)\simeq\frac{2E_{1}(4\Omega)^{n/2-1} \Gamma\left(n/2\right)}{n!}e^{-\mu
E_1 t}
\end{equation}
which, for $n\gg 1$, coincides with that given by Eq. (\ref{lefttail}).

\begin{figure}[ht]
\includegraphics[width=7.5cm,clip=]{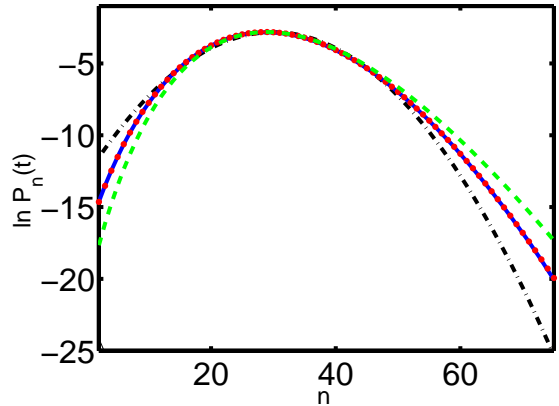}
\caption{(Color online) The natural logarithms of the analytical result
(\ref{probapprox}) for the quasi-stationary distribution (the dots), of the
distribution obtained by a numerical solution of the (truncated) master equation
(\ref{mastereq}) (the solid line), of the stationary solution (\ref{steadyFP})
of the Fokker-Planck equation (the dashed line), and of the gaussian
distribution (\ref{centerprob}) (the dashed-dotted line), for $\Omega=30$ and
$\mu E_1 t\ll 1$.} \label{probmaster}
\end{figure}

Figure \ref{probmaster} compares our analytical result (\ref{probapprox}) with
(i) a numerical solution of the (truncated) master equation (\ref{mastereq})
with $(d/dt)P_n(t)$ replaced by zeros and $P_0=0$, (ii) the prediction from the
Fokker-Planck equation for this problem [Eq.~(\ref{steadyFP}) of Appendix B],
and (iii) the gaussian distribution (\ref{centerprob}) for $\mu E_1 t \ll 1$. In
the central part all the distributions coincide. The Fokker-Planck approximation
strongly underpopulates the low-$n$ tail, and overpopulates the high-$n$ tail.
On the contrary, the gaussian approximation strongly overpopulates the low-$n$
tail and underpopulates the high-$n$ tail. Our analytical solution
(\ref{probapprox}) is essentially indistinguishable from the numerical result,
even at small values of $n$. Actually, it is in good agreement with the numerics
already for $\Omega={\cal O}(1)$, and the agreement improves further as $\Omega$
increases.

\begin{figure}[ht]
\includegraphics[width=7.5cm,clip=]{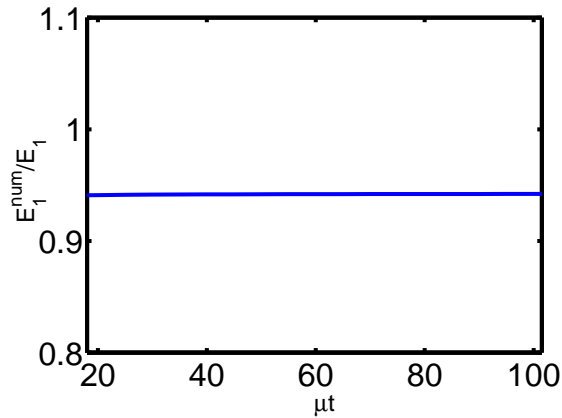}
\caption{Shown is the ratio of the numerical ground-state eigenvalue $E_1^{num}$
from Eq.~(\ref{gsnumerical}) and the approximate analytical value of $E_1$ from
Eq. (\ref{results}), for $\Omega=20$ and $n_0=100$. The deviation from $1$ is
about $5.6$ percent, that is within error ${\cal O}(1/\Omega)$.}
\label{eigenvalue}
\end{figure}

We also computed the ground-state eigenvalue by solving Eq.~(\ref{partdiffeq})
numerically with the boundary conditions $G(1,t)=1$, $\partial_x G(-1,t)=0$ and
the initial condition $G(x,t=0)=x^{n_0}$. At times $\mu t \gg 1/\Omega$, the
numerical ground-state eigenvalue $E_{1}^{num}$ can be found from the following
expression:
\begin{equation}\label{gsnumerical}
E_{1}^{num}=-\frac{1}{\mu t}\ln\left[1-G^{num}(0, t)\right]\,,
\end{equation}
where $G^{num}(x,t)$ is the numerical solution for $G(x,t)$, and the result in
Eq.~(\ref{gsnumerical}) should be independent of time. A typical example is
shown in Fig.~\ref{eigenvalue}, and  a good agreement with the theoretical
prediction (\ref{results}) is observed.

\section{Statistics of the extinction times}
The quantity $P_0(t)$, given by Eq. (\ref{p0}), is the probability of extinction
at time $t$. The extinction probability density is $p(t)=dP_0(t)/dt$. Using
Eq.~(\ref{p0}), we obtain the exponential distribution of the extinction times:
\begin{equation}\label{ltdist}
p(t)\simeq \mu E_{1} e^{-\mu E_{1} t}\;\;\;\mbox{at}\;\;\;\lambda
t\gg 1\,.
\end{equation}
The average time to extinction is, therefore,
\begin{equation}\label{meanextime}
\bar{\tau} = \int_0^{\infty}t p(t)\,dt\simeq(\mu
E_{1})^{-1}=\frac{2\sqrt{\pi}}{\mu\Omega^{3/2}}e^{2\Omega(1-\ln 2)}\,.
\end{equation}
This is in full agreement with the result of Turner and Malek-Mansour
\cite{turner}, and in disagreement with the prediction from the Fokker-Planck
approximation, given by Eq.~(\ref{tauFP}), and with the prediction from the
gaussian approximation, given by Eq.~(\ref{tauFPgauss}), see Appendix B.

Figure \ref{numexact} compares the analytical result (\ref{p0}) for $P_{0}(t)$
with the extinction probability $P_0^{num}(t)=G^{num}(0,t)$ found by solving
Eq.~(\ref{partdiffeq}) numerically as described at the end of the previous
section, and a very good agreement is observed.

\begin{figure}
\includegraphics[width=7.5cm,clip=]{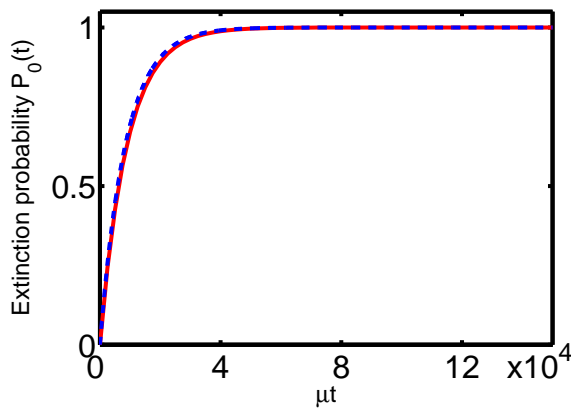}
\caption{(Color online) Shown are the extinction probability $P_{0}(t)$
[Eq.~(\ref{p0})] (the dashed line) and the numerical solution of
Eq.~(\ref{partdiffeq}) (the solid line) at $x=0$, for $\Omega=20$ and
$n_0=100$.} \label{numexact}
\end{figure}

\section{Final comments}
We calculated, at intermediate and long times,  the complete probability
distribution, including the quasi-stationary distribution of the long-lived
metastable state, and the extinction time statistics in a (non-single-step)
branching-annihilation reaction. To this end we employed the spectral method,
recently developed by the authors \cite{assaf,assaf1}. We also used this example
to illustrate how the ``lacking" boundary condition of the spectral method
emerges in the theory.

The spectral method reduces the problem of finding the statistics to that of
finding the ground-state eigenvalue and eigenfunction of a linear differential
operator emerging from the generating function formalism. The quasi-stationary
distribution that we have calculated analytically is in excellent agreement with
numerics. The two widely used ``rival" approximations: the Fokker-Planck
approximation and its reduced version, the gaussian approximation, perform well
only in the peak region of the quasi-stationary distribution. They both fail in
the tails of the distribution and, as a result, cause exponentially large errors
in the estimates of the mean extinction time.

It is worth reiterating that, for \textit{single}-step birth-death systems, the
quasi-stationary distribution can be found directly from a recursion equation
for $P_n$, obtained by putting $P_0=0$, assuming a zero flux into the empty
state, and replacing $(d/dt)P_n(t)$ by zeros in Eq. (\ref{mastereq}), see
\textit{e.g.} \cite{oppenheim}. For \textit{multi}-step systems such recursion
equations are not generally soluble analytically.

In conclusion, the spectral method is a powerful tool for calculating the
quasi-stationary distributions and extinction time statistics of a host of
multi-step birth-death processes possessing a metastable state and an absorbing
state.

\begin{acknowledgments}
We are very grateful to Alex Kamenev, Vlad Elgart and Lev Tsimring
for useful discussions. Our work was supported by the Israel Science
Foundation (grant No. 107/05) and by the German-Israel Foundation
for Scientific Research and Development (Grant I-795-166.10/2003).
\end{acknowledgments}

\section*{Appendix A}
\renewcommand{\theequation}{A\arabic{equation}}
\setcounter{equation}{0} Here we will derive the result given by
Eq.~(\ref{phibulkapprox}). First, we calculate the inner integral
\begin{equation}\label{a1}
I_1=\int_{-1}^{\infty}(1+r)^{2\Omega}e^{-2\Omega r}dr=
\left(\frac{e}{2\Omega}\right)^{2\Omega}\Gamma(2\Omega)\,.
\end{equation}
As $\Omega\gg 1$, we can use the large-argument asymptotics of the
gamma-function and obtain
\begin{equation}\label{a2}
I_1\simeq \sqrt{\frac{\pi}{\Omega}}\,.
\end{equation}
Second, at $\Omega \gg 1$, the integral
\begin{equation}\label{a3}
I_2=\int_{0}^{x}\frac{e^{2\Omega s}}{(1+s)^{2\Omega}}\,ds\equiv\int_0^x
e^{2\Omega \,\Upsilon(s)}\,ds
\end{equation}
where $\Upsilon(s)=s-\ln(1+s)$, receives the largest contribution in the
vicinity of $s=x$ (remember that $1-x\ll 1$). Then, expanding $\Upsilon(s)$
around $s=x$, we obtain
\begin{equation}\label{h(s)}
    \Upsilon(s)=x-\ln(1+x)+\frac{x}{1+x} (s-x)+\dots\,.
\end{equation}
Extending the lower integration limit to $-\infty$ and evaluating the remaining
elementary integral we obtain, in the leading order,
\begin{equation}\label{I2}
    I_2\simeq \frac{e^{2\Omega x}}{\Omega(1+x)^{2\Omega}}\,.
\end{equation}

\section*{Appendix B}
\renewcommand{\theequation}{B\arabic{equation}}
\setcounter{equation}{0} What are the predictions from the Fokker-Planck (FP)
approximation for the quasi-stationary distribution and the mean extinction time
of the branching-annihilation problem? The FP description introduces an (in
general, uncontrolled) approximation into the exact master equation
(\ref{mastereq}) by assuming $n\gg1$ and treating the discrete variable $n$ as a
continuum variable. The FP equation can be obtained from Eq.~(\ref{mastereq}) by
a Kramers-Moyal ``system size expansion" \cite{gardiner,vankampen} (in our case,
expansion in the small parameter $\Omega^{-1}\ll1$). Using this prescription, we
obtain after some algebra
\begin{eqnarray}
\frac{\partial P(n,t)}{\partial t} &=&
\frac{\mu}{2}\left\{-\frac{\partial}{\partial
n}\left[2n(\Omega-n)P(n,t)\right]\right.\nonumber\\
&+&\left.\frac{1}{2}\frac{\partial^2}{\partial
n^2}\left[2n(2n+\Omega)P(n,t)\right]\right\}\,. \label{FP}
\end{eqnarray}
The quasi-stationary distribution of the metastable state corresponds to the
(zero-flux) steady state solution of the FP equation. In the leading order in
$1/\Omega$ we obtain
\begin{equation}\label{steadyFP}
P_{st}(n)\simeq(3 \pi \Omega)^{-1/2}\, e^{\Omega-n+\frac{3}{2} \Omega \ln
\frac{2 n+\Omega}{3\Omega}}\,,
\end{equation}
where only the central (gaussian) part of the distribution contributes to the
normalization. In fact, the distribution~(\ref{steadyFP}) is accurate only in
the peak region, $|n-\Omega|\ll\Omega$ (see Section IV), where it reduces to a
gaussian distribution with mean $\Omega$ and variance $3\Omega/2$:
\begin{equation}\label{gauss}
P_{gauss} (n) = (3 \pi \Omega)^{-1/2}\,e^{-\frac{(n-\Omega)^2}{3\Omega}}\,.
\end{equation}
Were the FP equation (\ref{FP}) valid for all $n$, one could use it to find the
mean time to extinction $\tau_{FP}$ (conditional on non-extinction prior to
reaching the quasi-stationary state) by a standard calculation, see
\textit{e.g.} Ref. \cite{gardiner}. This calculation would yield
\begin{equation}\label{tauFPcalc}
\tau_{FP}\simeq 2\!\int_0^{\Omega}e^{n}\!
\left(1\!+\!\frac{2n}{\Omega}\right)^{-3\Omega/2}\!dn\!\int_{n}^{\infty}\frac{e^{-k}
\left(1\!+\!\frac{2k}{\Omega}\right)^{3\Omega/2}}{\mu
k(2k+\Omega)}dk.
\end{equation}
As $\Omega\gg 1$, the inner integral receives its main contribution from the
vicinity of $k=\Omega$.  The outer integral receives its main contribution from
the vicinity of $n=0$.  Therefore, one can use the saddle point method for the
inner integral and a Taylor expansion for the outer one. The result is
\begin{equation}\label{tauFP}
\tau_{FP}\simeq\sqrt{\frac{\pi}{3}}\frac{1}{\mu
\Omega^{3/2}}e^{\Omega\left(\frac{3}{2}\ln 3-1\right)}\,.
\end{equation}
Comparing it with Eq.~(\ref{meanextime}), one can see that the FP approximation
gives a poor estimate of the mean extinction time, as it introduces an
exponentially large error.

The central (gaussian) part of the quasi-stationary distribution,
$|n-\Omega|\ll\Omega$ (\ref{gauss}), can be \textit{correctly} obtained by
keeping only leading order terms, in the small parameter $|n-\Omega|/\Omega\ll
1$, in the FP-equation:
\begin{eqnarray}
\frac{\partial P(n,t)}{\partial t} &=&
\frac{\mu}{2}\left\{-\frac{\partial}{\partial
n}\left[2\Omega(\Omega-n)P(n,t)\right]\right.\nonumber\\
&+&\left.\frac{1}{2}\frac{\partial^2}{\partial
n^2}\left[6\Omega^2P(n,t)\right]\right\}\,. \label{FPgauss}
\end{eqnarray}
Indeed, the zero-flux steady state solution of this equation yields the gaussian
distribution (\ref{gauss}).

Finally, what would be the prediction for the mean extinction time from the
\textit{reduced} FP description, that is the one in terms of
Eq.~(\ref{FPgauss})? Here one would obtain
\begin{equation}\label{tauFPgauss}
\tau_{gauss}\simeq2\int_0^{\Omega}e^{\frac{n^2}{3\Omega}-\frac{2n}{3}}dn\int_{n}^{\infty}\frac{e^{\frac{2k}{3}-\frac{k^2}{3\Omega}}}{3\mu
\Omega^2}dk\simeq \frac{\sqrt{3\pi}}{\mu
\Omega^{3/2}}e^{\frac{\Omega}{3}}\,,
\end{equation}
which again gives an exponentially large error as compared with the accurate
result (\ref{meanextime}).

\section*{Appendix C}
\renewcommand{\theequation}{C\arabic{equation}}
\setcounter{equation}{0} Here we calculate the $n$-th derivative of
$\varphi_b(x)$, given by Eq. (\ref{phibulkapp}), at $x=0$. The first derivative
is
\begin{eqnarray}
\varphi_{b}^{\prime}(x)&=&-\frac{E_{1}}{\Omega}\left(\frac{e}{2\Omega}\right)^{2\Omega}
\frac{e^{2\Omega
x}}{(1+x)^{2\Omega}}\nonumber\\
&\times&\left\{\Gamma[2\Omega+1]-\Gamma[2\Omega+1,2\Omega(1+x)]\right\}.
\label{b1}
\end{eqnarray}
Let us introduce two auxiliary functions:
\begin{eqnarray}
f(x)&=&\frac{\left\{\Gamma[2\Omega+1]-\Gamma[2\Omega+1,2\Omega(1+x)]\right\}}
{(1+x)^{2\Omega}}\,,\nonumber\\
h(x)&=& e^{2\Omega x}\,.\label{b2}
\end{eqnarray}
Using Eqs.~(\ref{b1}) and (\ref{b2}), we can write the $n$-th derivative of
$\varphi_b(x)$ [that is, the $(n-1)$-th derivative of $\varphi_b^{\prime}(x)$]
at $x=0$ as
\begin{eqnarray}
\left.\frac{d^n\varphi_{b}(x)}{dx^n} \right|_{x=0}&=
&-\frac{E_{1}}{\Omega}\left(\frac{e}{2\Omega}\right)^{2\Omega}
\,\,\sum_{k=0}^{n-1}\frac{(n-1)!}{k!(n-k-1)!}\nonumber\\
&\times&\left.f^{(k)}(x)\right|_{x=0}\left.h^{(n-1-k)}(x)\right|_{x=0}\,,
\label{b3}
\end{eqnarray}
where $f^{(k)}(x)$ is the $k$-th derivative of $f(x)$, and the same notation is
used for $h(x)$.

After some algebra, we find that the $k$-th derivative ($k\geq 1$) of $f(x)$ at
$x=0$ is \cite{Abramowitz,Gradstein}
\begin{equation} \label{b4}
\left.\frac{d^k f(x)}{dx^k}\right|_{x=0}=(-1)^k 2\Omega
\left[\Gamma(2\Omega+k)-\Gamma(2\Omega+k,2\Omega)\right]\,.
\end{equation}
Now, the $k$-th derivative of $h(x)$ at $x=0$ is
\begin{equation} \label{b5}
\left.\frac{d^k h(x)}{dx^k}\right|_{x=0}=(2\Omega)^{k}\,.
\end{equation}
Using Eqs. (\ref{prob}), (\ref{b3}), (\ref{b4}) and (\ref{b5}), we obtain for
$n\geq 1$:
\begin{eqnarray}
P_{n}(t)&=&e^{-\mu E_1
t}\,\,\frac{2E_{1}}{n}\left(\frac{e}{2\Omega}\right)^{2\Omega}\,\,
\sum_{k=0}^{n-1}\frac{(-1)^k(2\Omega)^{n-k-1}}{k!(n-k-1)!}\nonumber\\
&\times&\left[\Gamma(2\Omega+k)-\Gamma(2\Omega+k,2\Omega)\right]\,. \label{b6}
\end{eqnarray}
Actually, for $n\!=\!1$ one has $$P_1(t)\simeq
(\pi/\Omega)^{1/2}\,E_1\left[1+{\cal O}\left(\Omega^{-1/2}\right)\right],$$ and
the sub-leading term ${\cal O}(\Omega^{-1/2})$ has been neglected in Eq.
(\ref{b6}). Finally, using Eq.~(\ref{gammaexpansion}) and changing the order of
summation in Eq.~(\ref{b6}), we obtain the following result for $n\geq 1$:
\begin{equation}
\label{b7}
P_{n}(t)\!=\!\frac{2E_{1}}{n}\!
\frac{(2\Omega)^{n\!-\!1}e^{2\Omega}\Gamma(2\Omega)}{\Gamma(2\Omega+n)}\,_1\!
F\!_1(2\Omega,n\!+\!2\Omega,-\!2\Omega)e^{-\mu E_1 t},
\end{equation}
where $_1\! F\!_1(a,b,x)$ is the Kummer confluent hypergeometric
function \cite{Abramowitz}. To avoid excess of accuracy, we need to
work with the large-$\Omega$ asymptotics of this result, see Section
IV.

\end{document}